\documentclass{mem}
\usepackage{natbib}\usepackage{txfonts}\usepackage{balance}
\usepackage{graphicx}
\idline{75}{282}
\begin{document}
\def\teff{$T\rm_{eff }$}
\def\kms{$\mathrm {km s}^{-1}$}

\title{
The most powerful explosions in the Universe: genesis and evolution of Supernova and Gamma-Ray Burst Italian programs at ESO
}

   \subtitle{}

\author{
E.  Pian\inst{1,2,3} 
          }

  \offprints{E.  Pian}

\institute{
Istituto Nazionale di Astrofisica --
Osservatorio Astronomico di Trieste, Via Tiepolo 11,
I-34143 Trieste, Italy
\and
Scuola Normale Superiore di Pisa,
Piazza dei Cavalieri 7,
56126 Pisa, Italy
\email{elena.pian@sns.it}
\and
INFN-Sezione di Pisa,
Largo Pontecorvo 3,
56127 Pisa, Italy
}

\authorrunning{Pian}

\titlerunning{Supernova and Gamma-Ray Burst at ESO}

\abstract{
The Italian communities engaged in Gamma-Ray Burst (GRB) and supernova research have been using actively the ESO telescopes and have contributed to improve and refine the observing techniques and even to guide the characteristics and performances of the instruments that were developed.  Members of these two communities have recently found ground for a close collaboration on the powerful supernovae that underlie some GRBs.  I will review the programs that have led to some  important discoveries and milestones on thermonuclear and core-collapse supernovae and on GRBs.
}
\maketitle{}

\section{Introduction}

Supernovae and Gamma-Ray Bursts (GRB) exhibit similar energy outputs, comparable to the binding energy of a compact star, $\sim$10$^{53}$ erg, that they release  in various forms, including electromagnetic radiation from radio to gamma-rays.   There is evidence that both explosion types are non-spherical, although the degree of asymmetry is very different, and its effect on the observed radiation is magnified by the presence of relativistic conditions, which are absent in thermonuclear supernovae, occasionally present and mild in core-collapse   supernovae and extreme in GRBs.  The evolution of the progenitor star influences the explosion, and its observable aftermath may depend on intrinsic conditions, but also  on the  circumstellar medium and on the viewing angle.  This generates a  highly diversified range of observed properties of the explosion remnants, that hold the key for the understanding of these phenomena.   Selected results obtained so far on these sources with observations at ESO telescopes by Italian scientists are here reviewed.

\section{Supernova programs at ESO}

Five years after Italy had joined ESO, Supernova 1987A went off in the Large Magellanic Cloud, and drove a flurry of observations and theoretical studies that contributed to boost the interest in supernova research in Italy and to establish systematic observing programs.  Among these, two ESO Key Programs  led by the group at Padua Astronomical Observatory, and involving the 3.6m, NTT, 2.2m, and Dutch telescopes,  were started in the 1990s.  These  contributed critically  to increase the statistics of well monitored supernovae and can be considered the precursors of ESO observing projects on unpredictable targets, i.e. sources whose coordinates -- because of their explosive nature -- cannot be known a priori.   Thereafter, supernova programs have been entertained by the Padua team at ESO regularly and have had an impact on international supernova research, including the characterization of ``branch-normal Type Ia supernovae" (Patat et al. 1996), of sub-luminous Ia supernovae (Turatto et al. 1996), and the contribution to  work on cosmology (Hamuy et al. 1996; Perlmutter et al. 1998) that eventually led to the discovery of the accelerated expansion of the Universe (Riess et al. 1998; Schmidt et al. 1998; Perlmutter et al. 1999).

This detailed work on individual Type Ia supernovae\footnote{a.k.a.   thermonuclear supernovae, to be distinguished from core-collapse supernovae} has underlined the remarkable diversity in these sources,  that were traditionally considered rather homogeneous -- at least photometrically -- and as such used as standard candles.   Supernovae Ia have been known for about 20 years not to be uniform after all, because their maximum luminosities span a magnitude or so.  The relationship reported by Phillips (1993) between the maximum luminosity and the rate of light decay in the B band after maximum  restores the reliability of Type Ia supernovae as standard candles, but, again, not all of them comply with this relationship.   A third parameter seems to be necessary to describe the full range of Ia supernovae properties, and this has been identified as the velocity gradient, i.e. the rate at which the photospheric velocity, estimated from the blue-shift of the  SiII 6355\AA\ line,  decreases with time. 

In a seminal paper by S. Benetti and collaborators (2005), three sub-classes of objects are distinguished in a sample of 26 well monitored Ia supernovae by their location in the space of photospheric velocity decline rate and maximum luminosity.  This segregation is suggested to be related to the type of explosion, i.e. deflagration versus detonation, or a transition between the two (delayed detonation).    However, subsequent work on this same sample of Ia supernovae, based on both early and late photometry and spectra, showed that almost all explosions are produced by the burning of a white dwarf star of rather constant mass, $1.05 \pm 0.09$ M$_\odot$, the ones that synthesize more nuclear statistical equilibrium elements ($^{56}$Ni, $^{58}$Ni, $^{54}$Fe) producing less intermediate mass elements (e.g. silicon), and viceversa, so that the sum of their masses is constant with relatively small dispersion.  This result is summarized in the flipped-Z-shaped diagram\footnote{Sometimes called ``Zorro" diagram} of mass versus $\Delta m_{15}(B)$ discussed in Mazzali et al. (2007a) and it suggests that the explosion mechanism is the same in all cases (most probably delayed detonation), with the number of ignition spots determining the final nucleosynthesis.  

Later on, it was suggested that the differences  within the sub-classes of Ia supernovae based on photospheric velocity decline rates 
can be determined by a different viewing angle, implying that the explosion is asymmetric (Maeda et al. 2010), a result  confirmed also by ESO spectropolarimetry (Maund et al. 2010).

\section{Gamma-Ray Bursts programs at ESO}

Until early 1997, GRBs were one of the biggest outstanding enigmas of astrophysics: the lack of detected counterparts at lower energies than the gamma-rays was preventing any firm conclusion about their nature
and even about their distance scale, conjectures ranging from galactic to cosmological scenarios (Lamb 1997; Paczy\'nski 1997).  GRB970228 was localized with arc-minute precision by the  Italian-Dutch {\it BeppoSAX}  satellite, which made possible the detection of its X-ray afterglow,  the first GRB counterpart ever detected (Costa et al. 1997).  The afterglow era had started.  The accurate localization by the {\it BeppoSAX}  instruments of the GRB and its afterglow allowed also the unambiguous identification of the counterpart at optical wavelengths (Van Paradijs et al. 1997).  The NTT observations of the GRB field, made  two  weeks after the GRB by the team led by the late Jan Van Paradijs,  that included all the Italian {\it BeppoSAX} group,  identified the host galaxy of the GRB, later confirmed with HST observations, which pointed for the first time in a direct and unambiguous way to the extragalactic origin of GRBs.   

Two years later, the Italian team, among others (Stanek et al. 1999; Beuermann et al. 1999), had a relevant role in inferring the presence of relativistic jets in GRBs from the steepening of the optical light curve of the afterglow of GRB990510, observed with the ESO  NTT, VLT and 3.6m telescope (Israel et al. 1999).  Consistent with jetted emission, significant optical linear polarization at the 1.6-1.7\% level was independently detected in the same GRB afterglow by two teams, one of which was under Italian leadership (Covino et al. 1999; Wijers et al. 1999).

GRB990123 confirmed a prediction that GRBs, besides exhibiting multi-wavelength afterglows, have prompt {\it simultaneous} very bright optical counterparts, called ``optical flashes", likely produced by  internal  or by reverse shocks,  while afterglows are rather due to the  interaction of the GRB blast-wave with the circumstellar medium.  These optical flashes can be extremely bright (in GRB990123, the first one detected, it peaked at $V \sim 9$, Akerlof et al. 1999), but they are also dramatically ephemeral, so that only small, flexible robotic telescopes can observe and detect them.  The record holder is 
GRB080913B,  also called ``naked-eye GRB", because its optical flash reached the 5th magnitude.  Its brightness made it suitable for very fast photometry with the Italian telescope REM at La Silla, equipped with the TORTORA instrument, that detected extremely rapid (time-scale of few seconds) optical variations correlated with the gamma-rays variations (Racusin et al. 2008).   The bright optical afterglow was also the target of VLT UVES observations in Rapid Response Mode, that started at 8.5 minutes after the GRB onset, when the magnitude was R $\sim$12. 
In the UVES spectrum were detected absorption features belonging to the main system at z = 0.937 and divided in at least six components spanning a total velocity range of 100 km s$^{-1}$.  These are estimated to be at distances between 2 and 6 kpc from the GRB site. 
Strong and variable Fe II fine structure lines were also  observed, whose optical depth variation correlated with the UV/optical flux decrease, thus proving  that the excitation of the observed fine structure lines is due to
pumping by the GRB UV photons (D'Elia et al. 2009).  

The host galaxies of GRBs bear the signatures of the stellar populations in which the explosion has taken place, and may therefore provide a way to characterize the class of progenitors of these powerful events, through the study of their basic properties, like luminosity, stellar mass, intrinsic absorption, star-formation rate, metallicity.   The Italian astronomers have been responsible for many studies of the host galaxies of GRBs with ESO telescopes.  Recently, a pilot study of GRB hosts with the newly deployed spectrograph X-Shooter, in the framework of the Italian-French Guaranteed Time program,  underlined the potential of this new instrument in the accurate study of the metallicity of GRB hosts.  The numerous lines that X-Shooter can detect  make possible the application of a number of metallicity indicators and extend  the investigation to the ``redshift desert", i.e. the interval $1 < z < 2$, for which the access to the optical range alone is not completely effective  for host galaxy emission line detection  (Piranomonte et al. 2011).

\section{Gamma-Ray Bursts meet supernovae}

One year after the outbreak of the GRB-afterglow era, an energetic Type Ic supernova\footnote{i.e., a core-collapse supernova whose progenitor has lost both its hydrogen and helium envelopes, to be distinguished from Type Ib supernova, with helium but no hydrogen and from Type II supernova, where both hydrogen and helium are kept before explosion.}, SN~1998bw, at 35 Mpc was detected and identified in the error box of the GRB980425, that had been detected by the {\it CGRO} BATSE and {\it BeppoSAX} GRBM and localized to arc-minute precision by the {\it BeppoSAX}  Wide Field Cameras (Soffitta et al. 1998; Galama et al 1998; Pian et al. 2000).   That no  optical afterglow was detected for this GRB, with the classical power-law time decay that astronomers had become accustomed to identify during one year of afterglow studies, came to a surprise\footnote{GRB980425 exhibited no X-ray afterglow either, but its  X-ray counterpart, which was identified with the X-ray emission from SN~1998bw, decreased by a factor of 2 in six months, and  further decayed by a factor of 10 in the following 4 years (Pian et al. 2000; Kouveliotou et al. 2004).}.  

SN~1998bw, whose explosion time was consistent with the GRB trigger time within $\sim$1 day, was exceptionally bright for a core-collapse supernova, having reached at maximum a luminosity comparable to that of a thermonuclear supernova (the synthesized radioactive $^{56}$Ni, estimated from radiation transport models, is $\sim$0.4 M$_\odot$), and even more importantly, it exhibited initial photospheric velocities of up to 30000 km/s, that implied a kinetic energy of $5 \times 10^{52}$ erg, more than an order of magnitude larger than the  typical $10^{51}$ erg (Mazzali et al. 2006a).  

Later on, supernova features have been identified  up  to $z \sim 1$ in a large fraction of optical afterglows of long GRBs\footnote{The  majority of GRBs have durations longer than $\sim$2 s (Kouveliotou et al. 1993) and  are thus defined as long GRBs, as opposed to short or sub-second GRBs.  Supernovae have been so far detected only for long GRBs.}  with well monitored  light curves and accurately measured optical spectra as observed with ESO telescopes by European teams, including or led by Italian scientists (e.g., Hjorth et al. 2003; Della Valle et al. 2003;  Malesani et al. 2004; Ferrero et al. 2006; Pian et al. 2006; Sparre et al. 2011; Bufano et al. 2012; Melandri et al. 2012).  
The GRB-supernova connection is best studied at  $z \lesssim 0.3$, because the low distance allows  spectroscopic monitoring of good signal-to-noise ratio and a full characterization of the supernova (see Figure~1).  These supernovae are all of Type Ic, i.e. their  spectra  lack hydrogen and helium lines, and are therefore the explosion of highly stripped massive stars, that have entered the Wolf-Rayet phase.   

Two of the GRBs associated with low-redshift well monitored supernovae are actually X-ray Flashes (XRFs), i.e. GRBs with softer spectra (i.e. peaking in the X-rays rather than the gamma-rays).   Not only the synthesized $^{56}$Ni and kinetic energies of GRB-supernovae are larger than in normal core-collapse supernovae\footnote{The properties of XRF-associated supernovae are intermediate between those of GRB-supernovae and normal Ic supernovae, see Figure~1.}, but also their ejecta masses  (Mazzali et al. 2006a).   This points to progenitor main sequence masses of at least 20 M$_\odot$ and, for the most massive GRB-supernovae,  40 M$_\odot$ (Mazzali et al. 2006a;b; Bufano et al. 2012).  The latter is considered to be the upper limit for the main sequence mass of a star undergoing supernova.  Beyond this mass, the core should collapse directly to a black hole (e.g. Heger et al. 2003), perhaps after going through  a brief (few seconds?) neutron star phase.  In this case, if a GRB is formed, it will not be accompanied by a supernova, which may explain  the lack of supernova detection, down to very deep photometric limits obtained with the VLT and HST, in  the $z \sim 0.1$ GRB060614 and GRB060505 (Della Valle et al. 2006; Fynbo et al. 2006; Gal-Yam et al. 2006).  

Finally, while GRB and XRF supernovae have maximum brightnesses in a range of 1.5 magnitudes
(Figure~1), the energy outputs of their high energy events span four orders of magnitude (Melandri
et al. 2012, and references therein).  This range is even wider if we include the higher redshift GRBs, whose promptly
emitted gamma-ray energy is $10^{52}$ to $10^{54}$ erg.  The wide interval of GRB gamma- and
X-ray luminosities and the fact that local GRBs/XRFs are under-energetic point to the role of
the viewing angle, that influences dramatically the jet emission via special relativity effects.
While SN luminosities are much less affected by these, the explosion may be highly anisotropic, which will be reflected in the SN light curves and spectra, especially at late epochs ($\gtrsim$200 days after explosion), when the inner ejecta are exposed and nebular, forbidden emission lines are formed.  The profiles of these nebular lines, markedly those of oxygen and iron, have been studied with VLT FORS2 spectroscopy and they reveal details on the geometry of the explosion, on the degree of asymmetry and on the viewing angle (e.g. Mazzali et al. 2007b; Valenti et al. 2012).

The low redshifts at which GRB and XRF supernovae are best studied make also their close environments better targets for accurate  studies.
This is efficiently accomplished through space-resolved spectroscopy of a small region centered on the GRB site.  An example is the study of the $z = 0.49$ host galaxy of GRB091127, associated with SN2009nz (Cobb et al. 2010;  Berger et al. 2011), for which  X-Shooter Integral Field Unit spectroscopy evidenced, via the construction of two-dimensional velocity, velocity dispersion, and star-formation rate  maps, perturbed rotation kinematics and a star-formation rate enhancement consistent with the supernova position (Vergani et al. 2011).

\begin{figure*}[t!]
\resizebox{\hsize}{!}{\includegraphics[clip=true]{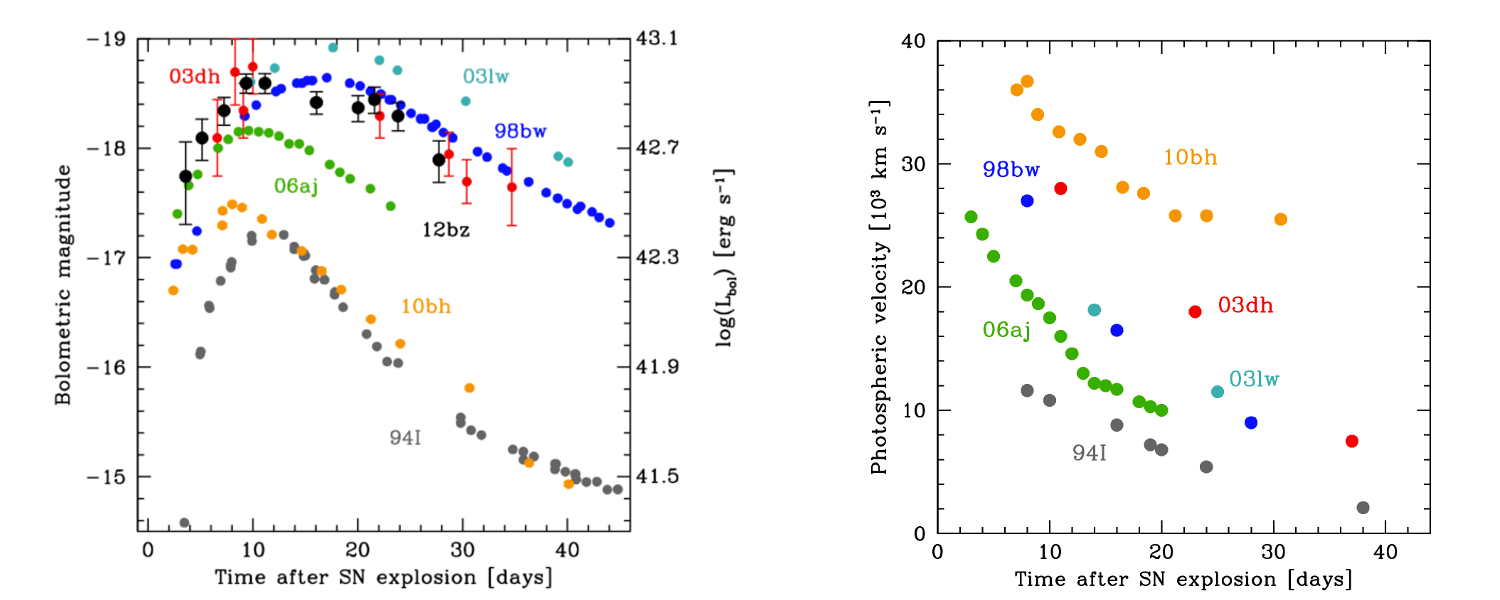}}
\caption{\footnotesize
{\bf (left)} Bolometric light curves of GRB and XRF supernovae at $z < 0.3$ (all observed with the ESO NTT and VLT) and of the ``normal"  (i.e., not accompanied by a GRB) Type Ic SN~1994I in rest frame (adapted from Melandri et al. 2012).  For GRB/XRF supernovae, the explosion time is assumed to coincide with the GRB/XRF  trigger time.  The explosion date of SN~1994I was obtained
from  light-curve models.  The supernovae associated with GRBs (GRB980425/SN~1998bw, 35 Mpc; GRB030329/SN~2003dh, $z = 0.168$;  GRB021203/SN~2003lw, $z = 0.106$; GRB120422A/SN~2012bz, $z = 0.283$) are the most luminous (synthesized radioactive $^{56}$Ni of  0.3-0.4 M$_\odot$) and are similar.   The XRF supernovae (XRF060218/SN~2006aj, $z = 0.033$ and XRF100316D/SN~2010bh, $z = 0.059$) are less luminous.  SN~2010bh is comparable in maximum luminosity to SN~1994I (10 Mpc), but faster in its rise  to peak.  This rapid rise time may indicate a strong asymmetry in the explosion.    
{\bf (right)}  Time evolution of the photospheric expansion velocities of  GRB and XRF supernovae evaluated from the width of the SiII 6355\AA\ spectral absorption line  at various epochs  (color coding as in left panel).   The large velocities, especially at early epochs, imply  high kinetic energies ($\gtrsim 10^{52}$ erg).  Note that SN~2010bh is the only supernova for which the width velocities were measured from the spectra and not evaluated with the aid of a model.  The difference between the two methods  amounts to about  20\%, and the main uncertainty of the former method is related to the possible blending of the SiII 6355\AA\  line with other transitions.
}
\label{eta}
\end{figure*}

\section{Conclusions}

Many supernova surveys currently active or coming on line will soon make the study of these explosions systematic and accurate and will at the same time guarantee large statistics (with many hundreds of supernovae detected per year)  and high level of detail in the study of individual cases of particular and peculiar interest.  These include the Palomar Transient Factory,  the Panoramic Survey Telescope \& Rapid Response System and the SkyMapper,  all counting also on ESO telescopes for follow-up observations.  Among these is the 
Public ESO Spectroscopic Survey of Transient Objects with the ESO NTT, led by S. Smartt and with a strong Italian participation.     Similarly,  the {\it Swift} satellite and other orbiting facilities for high energy astrophysics 
guarantee a rate of detection of GRBs and XRFs of about 100 events per year.  A fraction of these can be followed up within few minutes by the VLT, thanks to the Rapid Response Mode, that will secure, after prompt identification via acquisition photometry, timely spectroscopy  with the VLT UVES or X-Shooter.
Later on, associated supernova search and study of the environment will take place, assuming adequate observing time can be devoted to these programs.
Needless to say, besides clarifying existing problems, these observing programs  will open new ones and will bring along the serendipitous detection of many transients of unexpected nature, that may  create new interest and new research sub-branches in the two already very rich topics of supernovae and GRBs.  This will also help shape up the characteristics and performances of the future extremely large telescopes and of their instruments.

\begin{acknowledgements}
I am grateful to A. Antonelli, S. Benetti, S. Covino, V. D'Elia,  M. Della Valle, P. Ferrero, D. Malesani, N. Masetti, P. Mazzali, E. Palazzi  and M. Turatto  for their collaboration and for sharing with me their experiences and memories on  GRB and supernova observational activity at ESO in the last 30 years.  I would like to thank the organizer Vincenzo Mainieri for a very stimulating Symposium.  In the last 10 years, the supernova and GRB research programs have benefitted, among other sources of funding,  from the dedicated  European FP5 contracts for Research Training Networks  HPRN-CT-2002-00303  ({\it The Physics of Type Ia Supernovae}) and  HPRN-CT-2002-00294  ({\it Gamma-Ray Bursts: an Enigma and a Tool}),   from INAF PRIN contracts in 2006, 2009, 2011,  from PRIN MIUR 2005, and from ASI-INAF contracts I/023/05/0,  I/088/06/0, I/016/07/0, I/009/10/0.
\end{acknowledgements}

\bibliographystyle{aa}

\end{document}